\documentstyle[prb,aps,epsf]{revtex}

\tighten

\begin{document}

\twocolumn[\hsize\textwidth\columnwidth\hsize\csname @twocolumnfalse\endcsname

\title{Topology of amorphous tetrahedral semiconductors on intermediate
lengthscales}

\author{Normand Mousseau\cite{mousadd} and Laurent J. Lewis \cite{lewadd}}

\address{D{\'e}partement de physique et Groupe de recherche en physique et
technologie des couches minces (GCM), Universit{\'e} de Montr{\'e}al, Case
Postale 6128, Succursale Centre-Ville, Montr{\'e}al, Qu{\'e}bec, Canada, H3C 3J7}

\maketitle

\begin{center}
Submitted to Physical Review Letters
\end{center}

\begin{abstract}
Using the recently-proposed ``activation-relaxation technique'' for
optimizing complex structures, we develop a structural model appropriate to
{\it a}-GaAs which is almost free of odd-membered rings, i.e., wrong bonds,
and possesses an almost perfect coordination of four. The model is found to
be superior to structures obtained from much more computer-intensive
tight-binding or quantum molecular-dynamics simulations. For the elemental
system {\it a}-Si, where wrong bonds do not exist, the cost in elastic energy
for removing odd-membered rings is such that the traditional
continuous-random network is appropriate. Our study thus provides, for the
first time, direct information on the nature of intermediate-range topology
in amorphous tetrahedral semiconductors.
\end{abstract}
%\pacs{PACS: 61.43.Dq, 61.43.Bn, 02.70.Lq}

\vskip2pc]

\narrowtext

The structure of amorphous materials on short and intermediate lengthscales
remains largely unresolved in spite of decades of work on the
problem.\cite{zachariasen32,polk71,connell74,wooten85,lannin87,elliott89,yonezawa96}
While diffraction experiments can {\em in principle} provide the desired
information, they still lack the sensitivity necessary to distinguish between
various possible conformations of the structure. They must be backed up,
therefore, by accurate structural models in order to allow a meaningful
interpretation of the data. Models, however, are difficult to construct
because they require an accurate description of the interatomic forces, which
in turn limits the timescale over which simulations can be carried out. Since
glasses relax very slowly, this is an extremely serious problem that can only
be addressed through the use of judicious optimization methods.

Amorphous tetrahedral semiconductors, either elemental or compound, are
potential candidates for the fabrication of micro- and opto-electronic
devices. Advances in this area have however been hindered by the limited
understanding of their structure, and thus electronic properties, even at the
most elementary level. For instance, in the case of amorphous silicon ({\it
a}-Si), the average number of nearest-neighbors is still not known precisely,
even though evidence (experimental and theoretical) accumulates that it is
very close to 4. Thus, it would appear that the fundamental ``building
block'' of these materials is a tetrahedron --- a single atom and its four
nearest-neighbors arranged in a pattern that closely resembles that found in
the crystals. It is in the arrangement of those tetrahedra, i.e., on
intermediate (and larger) lengthscales, that differences between amorphous
and crystalline materials arise\cite{lannin87,elliott89} and on which,
therefore, attention should be focused.

The case of compound III-V materials, such as GaAs and InP, is more
problematic because the building blocks are not as clearly defined. Indeed,
the chemical identity of the atoms constituting the tetrahedron may vary:
While in the zinc-blende crystal each atom of a given species is surrounded
by four atoms of the other species, this is not necessarily so in the
amorphous material, because of disorder. Thus, there is a possibility that
``wrong bonds'' be present in the amorphous phase, i.e., bonds between like
atoms. This will result in some distortion of the tetrahedra, and therefore
on their relative arrangement. However, III-V compounds are partly ionic and
the cost in energy of wrong bonds is expected to be significant; the number
of them, therefore, will be determined by the competition between elastic
deformation and chemical disorder. There is currently no experimental method
for measuring this number in {\it a}-GaAs.\cite{udron91}

In this Letter, we aim at determining the optimal density of wrong bonds in
amorphous III-V semiconductors through a comparison with elemental materials,
where such defects do not exist. We illustrate our ideas by considering the
prototypical materials {\it a}-Si and {\it a}-GaAs. We note first that wrong
bonds are closely related to the presence of ``odd-membered rings'' in the
structure, i.e., closed paths between an atom and itself containing an odd
number of bonds. Evidently, only even-membered rings are present in
zinc-blende crystals. Also, it is always possible to ``decorate'' an
amorphous network containing only even rings with two types of atoms such
that there are no wrong bonds. Thus, we may formulate the question in the
following manner: are odd-membered rings energetically favorable, i.e.,
necessary to the stability of amorphous tetrahedral semiconductors, be they
elemental or compound? And is it possible to construct a tetrahedrally-bonded
amorphous model {\em without} odd-membered rings?

Anticipating our results, we find that it is indeed quite possible, albeit
difficult, to construct such an all-even-ring network, but the cost in
elastic-deformation energy required by this constraint is such that the
resulting model is {\em not} appropriate to elemental materials. In contrast,
a model that contains odd rings is found to be unappropriate for III-V
materials, because the cost in energy associated with wrong bonds largely
exceeds that of the elastic deformation. Thus, {\em odd-membered rings must
be present in {\it a}-Si and absent in {\it a}-GaAs.} This constitutes, to
our knowledge, the first evidence of a {\em topological difference} (i.e.,
without regards to chemical identity) between the two materials. We find,
also, that the presence (or absence) of odd-membered rings is extremely
difficult to extract from such quantities as the radial distribution function
or distribution of dihedral angles, and therefore realistic structural models
are a necessity for proper interpretation of experimental data.

In order to establish these results, we proceed as follows. Using the
``activation-relaxation'' technique\cite{barkema96} (see below), we construct
two ``generic'' continuous-random networks (CRN)\cite{zachariasen32} with
periodic boundary conditions: The first, CRN-A, possesses the same attributes
as an infinite Polk model\cite{polk71} as constructed by Wooten, Winer and
Weaire,\cite{wooten85} and thus contains odd-membered rings. (In fact, the
WWW algorithm consists in amorphizing a crystalline lattice through,
precisely, the introduction of odd rings). Our second model, CRN-B, was
constructed with the constraint that odd rings be absent, and thus
corresponds to an infinite Connell-Temkin\cite{connell74} computer model.
(The Connell-Temkin model is a hand-built, 238-atom, mechanical model with
free surfaces). Both CRN-A and CRN-B were then ``decorated'' with either Si
or Ga and As --- so that we have, altogether, 4 different structural models
--- and relaxed to their equilibrium state using conjugate gradients with
forces derived from tight-binding (TB) potentials.

It is important to note that there have been numerous attempts at simulating
the structure of {\it a}-Si using molecular dynamics (MD) and a variety of
interatomic potentials --- from empirical to fully {\em ab initio}. All lead
to a sizable proportion of odd-membered rings. For GaAs, for which there
exists no satisfactory empirical potential, calculations have been carried
out using either TB\cite{molteni94,seong96} or Car-Parrinello\cite{fois92}
(CP) MD. Again, here, odd-membered rings --- and therefore wrong bonds ---
are present, in an amount which appears to be quite large (10--12\%). Since
the period of time that can be simulated using such approaches is inevitably
short (of the order of 10--100 ps), there remains a possibility that the
large density of defects is the result of incomplete relaxation. In order to
resolve this point, a different, much more efficient, optimization method
must therefore be used to prepare the structural models.

The activation-relaxation Monte-Carlo technique (ART), recently proposed by
Barkema and Mousseau,\cite{barkema96}, provides a powerful and efficient way
of searching configuration space for a global minimum. Full details of the
method can be found in Ref.\ \onlinecite{barkema96}; it can be summarized as
follows: Starting with a random distribution of atoms, the system is first
relaxed to the nearest local minimum on the potential energy surface using
standard minimization techniques. It is then ``activated'' by pushing it to a
neighboring saddle point along a valley, i.e., a path of minimal energy. The
system is then relaxed, again, to the nearest local minimum, and the process
iterated until a global minimum is found.

The potential energy for the ART preparation of CRN-A was taken to be of the
Stillinger-Weber form\cite{stillinger85}, appropriate to Si, except that the
strength of the bond-bending term was increased by 50\% so as to provide a
better description of the material.\cite{barkema96} For CRN-B, the same
potential was used --- the lattice parameters of {\it c}-Si and {\it c}-GaAs
are nearly the same --- but a repulsive term between like atoms was added in
order to describe the chemical order:
   \begin{equation}
   E_{rep} = \sum_{<ij>} A_{ij} \epsilon \left[ 1 +
      \cos \left( \pi \frac{r_{ij}}{s_{ij}} \right) \right]
   \end{equation}
where the sum is over all pairs of atoms, $\epsilon$ is the Stillinger-Weber
energy parameter, $A_{ij}=1.2$ for like particles and zero otherwise, and
$s_{ij}=3.6$ \AA. Both CRN-A and CRN-B models contain 216 atoms in a cubic
cell with periodic boundary conditions. The ART relaxation took about three
days on a RS8000 workstation for CRN-A and about a week for CRN-B. This
should be contrasted with several weeks for a TB simulation of a 64-atom
system on a workstation or a CP simulation on a state-of-the-art parallel
supercomputer.

After reaching a stable configuration in the ART simulations, the two models
were made material-specific by further relaxing under TB potentials. {\em
Both} CRN-A and CRN-B models were thus relaxed with {\em both} the Vogl {\it
et al.} TB potential for Si\cite{vogl83} and the Molteni-Colombo-Miglio TB
model for GaAs.\cite{molteni94} In the case of model CRN-A-GaAs, it is
necessary to ``label'' the atoms. This was done by assigning identities at
random, then iteratively minimizing the number of wrong bonds by a
``label-switching'' process. This lead to a proportion of wrong bonds of
approximately 14\%, corresponding to the theoretical value for optimal
ordering on a Polk-type network.\cite{theye80}.

The energies of our four different TB-relaxed models are listed in Table
\ref{tab:energies}. It was observed that the topology of the models is little
affected by the final TB relaxation, indicating that the ART simulations
effectively took the systems very close to the global minimum.

Considering Si first, it is clear that the CRN-A model, which does contain
odd-membered rings, is a much more favorable structure for the material than
CRN-B, which has no such rings. The cost in elastic energy required to
eliminate odd-membered rings can be extracted from Table \ref{tab:energies}
as the difference in energy between CRN-A and CRN-B for Si, and comes out to
be about 0.08 eV/atom.

For GaAs, now, we find the opposite result: As indicated in Table
\ref{tab:energies}, it is much more interesting for this
\begin{table}[tb]
\caption{
Energy per atom of the two networks relaxed with TB. For GaAs, we also give
the results from the TB-MD simulations of Seong and Lewis (SL), Ref.\
\protect\onlinecite{seong96}.
}
\begin{tabular}{lccc}
Network & \multicolumn{2}{c}{TB parameters} &  \\
        &   Si    &  GaAs   &   GaAs (SL) \\ \hline
  CRN-A & -12.735 & -13.450 &             \\
  CRN-B & -12.659 & -13.561 &  -13.450    \\
Crystal & -13.084 & -13.802 &  -13.802    \\
\end{tabular}
\label{tab:energies}
\end{table}
\noindent material to adopt the CRN-B structure, free of odd-membered
rings and therefore wrong bonds,
than the CRN-A structure. Thus, the energy gained by the suppression of wrong
bonds {\em largely exceeds} the cost for elastically deforming the network to
eliminate odd-membered rings --- 0.08 ev/atom, as we have just seen.

The structural properties of the two optimal models --- CRN-A for Si and
CRN-B for GaAs --- are summarized in Table \ref{tab:coord}. Evidently, for
Si, we achieve an excellent-quality model, with an {\em almost perfect
coordination} of 4, sharply peaked at this value, i.e., a structure which,
while amorphous (see below), is {\em almost free of topological defects}.

\begin{figure}
\epsfxsize=7cm
\epsfbox{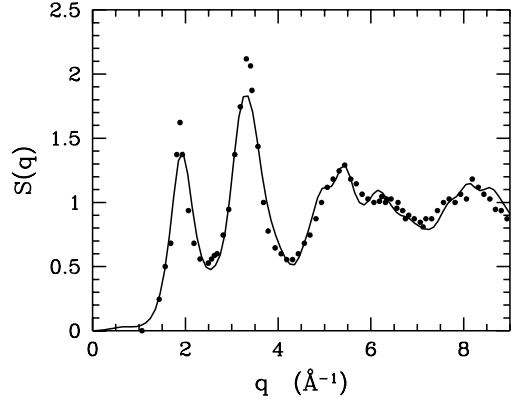}
\caption{
Total static structure factors for the CRN-B-GaAs model (full line) and from
experiment (dots, Ref.\ \protect\onlinecite{udron91}).
}
\label{fig:fss}
\end{figure}

Likewise, our CRN-B-GaAs model, with less than 4\% of wrong bonds and an
almost perfect coordination of 4, is by far the closest realization of a
perfect, infinite, Connell-Temkin-like CRN ever achieved. The present model
compares very favorably with experimental static-structure-factor
data,\cite{udron91} as demonstrated in Fig.\ \ref{fig:fss} as well as with
the results from TB- or CP-MD simulations, as can be seen in Tables
\ref{tab:energies} and \ref{tab:coord}. Using MD and exactly the {\em same}
TB model, Seong and Lewis\cite{seong96} (SL) obtained an amorphous model
whose energy lies significantly {\em above} that of the present model. (It si
not possible to compare energies from the CP\cite{fois92} and TB models). The
density of wrong bonds in the present model is also much lower than that
obtained in the TB- and CP-MD simulations. In addition, the distribution of
coordination numbers here is sharply peaked at about 4, while it is much
broader in the other models. These results suggest that the usual
melt-and-quench MD approach may not yield a fully-relaxed configuration,
while ART simulation approaches this point much more effectively. ART
relaxation under TB or first-principles forces is however presently not
available.

It is of interest to examine our models in terms of a possible structural
signature of the presence of odd-membered rings (or wrong-bonds), not based
on knowledge of the system at the atomic level, that would provide (e.g.,
from experiment) information on intermediate-range correlations. One
possibility is the radial distri-
\begin{table}
\caption{
Distribution of coordination numbers $Z$ and average value, nearest-neighbor
distance $r_{NN}$, and density of wrong bonds (WB) in our two optimized
models, compared with other models --- SL: TB-MD of Ref.\
\protect\onlinecite{seong96}; CP: CP-MD of Ref.\ \protect\onlinecite{fois92}.
For CRN-A, the density of wrong bonds is that of the GaAs-decorated lattice
as discussed in the text.
}
\begin{tabular}{lcccc}
        & CRN-A-Si & CRN-B-GaAs  & GaAs (SL) & GaAs (CP) \\ \hline
$Z=$ 3      & 0.037 & 0.051 & 0.242     &  0.219   \\
$Z=$ 4      & 0.963 & 0.944 & 0.598     &  0.781   \\
$Z=$ 5      & 0     & 0.005 & 0.129     &   0\\
$Z=$ 6      & 0     & 0     & 0.024     &   0\\
$Z=$ 7      & 0     & 0     & 0.007     &   0 \\
$<Z>$       & 3.963 & 3.954 & 3.94      &  3.83 \\
$r_{NN}$ (\AA) & 3.0 & 3.0 & 3.0 & 2.8 \\
WB (\%)     & 14.2 & 3.9 & 12.2 & 10.0 \\
\end{tabular}
\label{tab:coord}
\end{table}

\begin{figure}
\epsfxsize=7cm
\epsfbox{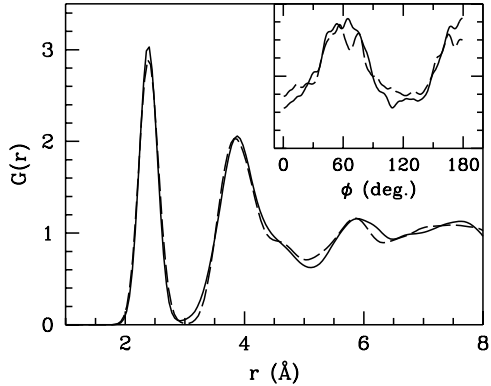}
\caption{
Radial distribution functions for the CRN-A-Si (dashed line) and the
CRN-B-GaAs (full line) models. The GaAs curve has been rescaled along $r$ so
as to match the first-neighbor peak of Si. The inset shows the distribution
of dihedral angles for the same two models.
}
\label{fig:fdr}
\end{figure}

\noindent
bution function, $G(r)$, which we show in Fig.\ \ref{fig:fdr} for our
 two best models, CRN-A-Si and CRN-B-GaAs. In
spite of the fact that these models differ markedly in their medium-range
topology, as we have seen, {\em the differences between the two curves are
clearly very small.} This is a bit of a surprise in that it contradicts the
current view that differences between the two structures should be clearly
visible beyond the second peak.\cite{gheorghiu85} In fact, it was argued by
Connell and Temkin\cite{connell74} that the signature of a CRN-B-type network
would be, in comparison to CRN-A, (i) a broader second peak, (ii) a deeper
minimum between the 2nd and 3rd peak as well as (iii) a shifted and (iv)
broader 3rd peak. This is not what we find here, as can be seen in Fig.\
\ref{fig:fdr}: although there are some differences --- not necessarily in
agreement with the above criteria --- they remain small and could easily be
within the accuracy of the models. Thus, $G(r)$ exhibits no clear signature
of the presence of odd-membered rings (even though the differences we observe
between the two structures are within experimental resolution).

We show, in the inset to Fig.\ \ref{fig:fdr}, the distribution of dihedral
angles (i.e., between second nearest-neighbor bonds) in our two models. In
the Connell-Temkin model\cite{connell74} (corresponding to our CRN-B),
staggered configurations ($\phi=60^{\circ}$) are found to be much more
probable than in the usual Polk CRN,\cite{polk71} which contains odd-membered
rings. Our models demonstrate that the differences are in fact very small and
again not significant. In particular both models seem to like staggered
configurations with comparable probability. Thus, the distribution of
dihedral angles is {\em not} a good measure of intermediate-range
correlations.

It follows from the above discussion that the nature of intermediate-range
correlations cannot easily be determined with current experimental
techniques, so that {\em atomic-level} models, which have the power to
resolve different possible structures yielding the same structure

\noindent factor on
the basis of total-energy calculations, remain the only resort for proper
interpretation of experimental data. Thus, while such a method as ``reverse
Monte Carlo'' (Ref.\ \onlinecite{rmc}) can provide almost perfect agreement
with diffraction data, it clearly cannot resolve two structures that yield
comparable structure factors because it does not include a total-energy
component.

In conclusion, using the newly-proposed activation-relaxation technique, we
have been able to construct a structural model appropriate to {\it a}-GaAs
that contains almost {\em no wrong bonds} (i.e., is free of odd-membered
rings) with an almost {\em perfect coordination} of four. This
Connell-Temkin-like model is found to be superior, from both structural and
energetic viewpoints, to structures obtained from TB and quantum MD
simulations, at a fraction of the computational cost. In contrast, in the
elemental system {\it a}-Si, where wrong bonds do not exist, the absence of
odd-membered rings costs elastic energy, so that the traditional Polk-like
CRN is appropriate. Our study thus provides, for the first time, direct and
solid information on the nature of intermediate-range correlations in
amorphous tetrahedral semiconductors, and, in particular, establishes clearly
that these materials cannot be described by a single topological model. It
underlines, moreover, the fact that state-of-the-art total-energy computer
models are essential for reliable interpretation of experimental diffraction
(and other) data. Finally, our study gives unambiguous evidence for the
ability of ART to yield the optimal structure of strongly-disordered systems,
thus opening a new way into ``complexity''.

% \acknowledgements

We acknowledge helpful discussions with G.T.\ Barkema and S.\ Roorda. This
work was supported by grants from the Natural Science and Engineering
Research Council (NSERC) of Canada and the ``Fonds pour la formation de
chercheurs et l'aide \`a la recherche'' of the Province of Qu\'ebec. Part of the
calculations were carried out at the ``Centre d'applications du calcul
parall\`ele de l'Universit\'e de Sherbrooke'' (CACPUS). We are grateful to the
``Services informatiques de l'Universit\'e de Montr\'eal'' for generous
allocations of computer resources.

\bibliographystyle{prsty}

\begin{thebibliography}{99}

\bibitem[(a)]{mousadd} Electronic mail address: mousseau@physcn.umontreal.ca

\bibitem[(b)]{lewadd} Electronic mail address: lewis@physcn.umontreal.ca

\bibitem{zachariasen32} W.H. Zachariasen, J. Am. Chem. Soc. {\bf 54}, 3841
(1932).

\bibitem{polk71} D.E. Polk, J. Non-Cryst. Sol. {\bf 5}, 365 (1971).

\bibitem{connell74} G.A.N. Connell and R.J. Temkin, Phys. Rev. B {\bf 9},
5323 (1974).

\bibitem{wooten85} F. Wooten, K. Winer, and D. Weaire, Phys. Rev. Lett. {\bf
54}, 1392 (1985); F. Wooten and D. Weaire, Solid State Physics {\bf 40}, 1
(1987).

\bibitem{lannin87} J.S. Lannin, J. Non-Cryst. Sol. {\bf 97\&98}, 39 (1987).

\bibitem{elliott89} S.R. Elliott, Advances in Phys. {\bf 38}, 1 (1989).

\bibitem{yonezawa96} F. Yonezawa, J. Non-Cryst. Sol. {\bf 198-200}, 503
(1996).

\bibitem{udron91} D. Udron, M.-L. Th\`eye, D. Raoux, A.-M. Flank, P. Lagarde,
and J.-P. Gaspard, J. Non-Cryst. Solids {\bf 137\&138}, 131 (1991).

\bibitem{barkema96} G.T. Barkema and N. Mousseau, Phys. Rev. Lett. in press
(code number LU6061).

\bibitem{molteni94} C. Molteni, L. Colombo, and L. Miglio, Phys. Rev. B {\bf
50}, 4371 (1994).

\bibitem{seong96} H. Seong and L. J. Lewis, Phys. Rev. B {\bf 53}, 4408
(1996).

\bibitem{fois92} E. Fois, A. Selloni, G. Pastore, Q.-M. Zhang, and R. Car,
Phys. Rev. B {\bf 45}, 13~378 (1992).

\bibitem{stillinger85} F.H. Stillinger and T.A. Weber, Phys. Rev. B {\bf 31},
5262 (1985).

\bibitem{vogl83} P. Vogl, H.P. Hjalmarson, and J.D. Dow, J. Phys. Chem.
Solids, {\bf 44}, 365 (1983).

\bibitem{theye80} M.-L. Th\`eye, A. Gheorghiu, and H. Launois, J. Phys. C:
Solid St. Phys. {\bf 13}, 6569 (1980).

\bibitem{gheorghiu85} A. Gheorghiu, K. Driss-Khodja, S. Fisson, M.-L. Th\`eye,
and J. Dixmier, J. de Phys. (France) {\bf C8}, 545 (1985).

\bibitem{rmc} D.A. Keen and R.L. Greevy, Nature {\bf 344}, 423 (1990); R.L.
McGreevy and L. Pusztai, Mol. Sim. {\bf 1}, 359 (1988).

\end{thebibliography}

\end{document}